\documentclass[summary]{ursi}

\usepackage{fontenc}
\usepackage{mathptmx}
\usepackage{geometry}
\usepackage{cite}
\usepackage{amsmath}
\usepackage{calrsfs}
\DeclareMathAlphabet{\pazocal}{OMS}{zplm}{m}{n}
\SetMathAlphabet\pazocal{bold}{OMS}{zplm}{bx}{n}
\usepackage{amssymb}
\usepackage{url}
\usepackage{graphicx}
\usepackage{color}
\usepackage{placeins}
\usepackage{float}
\usepackage{tabularx,colortbl}
\usepackage{ifthen}

\def\Rob#1{\textcolor{black}{#1}} 

\title{Multiport Network Modeling for Reconfigurable Intelligent Surfaces: Numerical Validation with a Full-Wave PEEC Simulator}

\author{Giuseppe Pettanice*\affref{ref1}, Marco Di Renzo\affref{ref2}, Sumin Jeong \affref{ref2}, Roberto Valentini \affref{ref1},  Piergiuseppe Di Marco \affref{ref1}, Fortunato Santucci\affref{ref1},  Daniele Romano\affref{ref1},
  and Giulio Antonini\affref{ref1}}

\affiliation{%
  \aff{ref1}{University of L’Aquila, 67100 L’Aquila, Italy (\{fortunato.santucci\} \{giulio.antonini\}@univaq.it)}
  \aff{ref2}{Universit\'e Paris-Saclay, CNRS, CentraleSup\'elec, Laboratoire des Signaux et Syst\`emes, 3 Rue Joliot-Curie, 91192 Gif-sur-Yvette, France. (marco.di-renzo@universite-paris-saclay.fr)}
}

\begin{document}

\maketitle

\begin{abstract}
  Reconfigurable Intelligent Surface \Rob{(RIS)} modeling and optimization are \Rob{a crucial steps} in developing the next generation of wireless communications. \Rob{To this aim, the availability of accurate electromagnetic (EM) models is of paramount important for the design of RIS-assisted communication links}. In this work, we \Rob{validate a widely-used analytical multiport network for RISs by means of a well-established full-wave numerical method based on the Partial Elements Equivalent Circuit (PEEC) approach}. Numerical results show good agreement between the two methods, \Rob{thus demonstrating i) the considered multiport network model being effective and ii) the  PEEC method being appropriate for EM modeling of RIS-assisted wireless links. }
\end{abstract}

\section{Introduction}
\label{sec:intro}
Reconfigurable intelligent surface (RIS) has been recognized as an innovative paradigm for electromagnetic (EM) wave manipulation, signal modulation, and smart radio environment reconfiguration \cite{Zapp10etal,DiRe20etal}. RISs are artificial metasurfaces composed of periodic or aperiodic subwavelength electric or magnetic resonators, which can \Rob{be controlled in order to induce} dynamic reflection/transmission amplitude and phase responses, thereby shaping reflected and transmitted wavefronts. \Rob{This allows for the reconfiguration of the scattering patterns according to the direction of the incident waves and the locations of the users.}

In order to analyze and optimize RIS-assisted wireless systems, communication models that consider the physics and EM characteristics of the RIS's scattering elements must be sufficiently realistic, accurate, and tractable. 
This calls for \Rob{EM} models that allow for an appropriate \Rob{description} of propagation \Rob{phenomena} and the mutual coupling among the radiating elements in the RIS. A circuit-based communication model for RIS-assisted wireless systems that is based on evaluating the mutual impedances between all the radiating elements (transmit/receive antennas, passive scatterers) is presented in \cite{Grad21DR}. A time domain macromodel of the RIS, including transmitters and receivers, has been introduced in \cite{Petta23}.

The scope of this work is twofold: i) to validate the analytical model of the RIS introduced in \cite{Grad21DR}, and ii) to prove the versatility of the \Rob{Partial Elements Equivalent Circuit (PEEC)} method for the EM modeling of communication channels. Specifically, we analyze the performance of RIS-aided channels by optimizing the tunable terminations of the RIS, and quantifying the obtained performance gains.

\section{System Model}
\label{sec:sysMOD}
\Rob{This section describes the two approaches employed to model RIS-aided channels: the numerical approach based on the PEEC method and the analytical framework based on multiport network theory.}
As stated in \cite{Petta23, Grad21DR}, the communication channel can be conceptualized as a multiport system. Accordingly, certain ports model the transmitters, while others model the receivers or scatterers. 

\Rob{The overall system can be characterized by the impedance matrix $\mathbf Z_{sys}$, which takes into account the interactions between transmitters, receivers, the RIS, as well as the presence of scattering objects in the environment.} This impedance matrix can be defined as
\begin{equation}
    {\mathbf Z_{sys}}=
    \left[
  \begin{array}{cccc}
    {\mathbf Z}_{TT} & {\mathbf Z}_{TS} & {\mathbf Z}_{TO} & {\mathbf Z}_{TR} \\
    {\mathbf Z}_{ST} & {\mathbf Z}_{SS} & {\mathbf Z}_{SO} & {\mathbf Z}_{SR} \\
    {\mathbf Z}_{OT} & {\mathbf Z}_{OS} & {\mathbf Z}_{OO} & {\mathbf Z}_{OR} \\
    {\mathbf Z}_{RT} & {\mathbf Z}_{RS} & {\mathbf Z}_{RO} &{\mathbf Z}_{RR} \\
  \end{array}
\right],
\end{equation}
where $\{T,R,S,O \}$ denote the transmitter, receiver, RIS, and the scattering objects in the environment, respectively.
\Rob{Moving from the knowledge of $\mathbf Z_{sys}$, the communication channel matrix $\mathbf{H}_{E2E}$  can be determined as described in \cite{HAS23}}
%
\begin{equation}\label{eq:H_E2E}
        \mathbf{H}_{E2E} = \mathbf Z_{RL} \left[ \mathbf Z_{ROT} - \mathbf Z_{ROS} \mathbf Z_{sca} \mathbf Z_{SOT} \right] \mathbf Z_{TG},
\end{equation}
where $\mathbf Z_{RL} = \left(\mathbb I_L + \mathbf Z_{RR} \mathbf Z_{L}^{-1} \right)^{-1}$, 
$\mathbf Z_{TG} = \left(\mathbb I_L + \mathbf Z_{TT} \mathbf Z_{G} \right)^{-1}$, and
$\mathbf Z_{sca} = \left(\mathbf Z_{SS} + \mathbf Z_{SOS} + \mathbf Z_{RIS} \right)^{-1}$. In particular, $\mathbb I_L$ is the identity matrix, $\mathbf Z_G$ and $\mathbf Z_L$ are the diagonal matrices containing the impedances of the voltage sources of the transmitters and the load impedances at the receiver; $\mathbf Z_{TT}$ and $\mathbf Z_{RR}$ are the matrices containing the self and mutual impedances at the transmitter and receiver;  $\mathbf Z_{ROT}$, $\mathbf Z_{ROS}$, $\mathbf Z_{SOS}$ and $\mathbf Z_{SOT}$ are the matrices containing the mutual impedances between different array elements, including the scattering objects in the environment.

\subsection{The PEEC Method}
\label{subsec:PEEC}
\Rob{The PEEC} method \Rob{relies on} a circuit-based model \cite{Ru17AJ} to represent EM phenomena related to transmitters, receivers, and scattering elements. \Rob{Specifically, the PEEC formulation is} based on the Electric Field Integral Equation (EFIE) and the continuity law for the electric current.
One of the key advantages of the PEEC method is its \Rob{capability} to \Rob{describe} the behavior of \Rob{complex} structures \Rob{by means of} standard circuit variables, namely node potentials and side electric currents.

\Rob{To obtain the PEEC representation}, the considered system, e.g.,  an antenna element, \Rob{is first} divided into smaller units (mesh), consisting of elementary volumes and surfaces. The electric currents are assumed to flow through the elementary volumes, while the electric charges are assumed to exist on the elementary surfaces of the mesh. \Rob{Then}, the coupling between the currents flowing in the volumes and the charges on the surfaces \Rob{are considered}. The magnetic interaction among the currents is described by partial inductances denoted as \Rob{$\mathbf{L}_p$}, while the electric interactions among the charges are represented by potential coefficients denoted as \Rob{$\mathbf{P}$}. Finally, the application of Kirchhoff laws to the PEEC equivalent circuit leads to the following Modified Nodal Analysis (MNA) representation in the frequency domain:
\begin{equation}\label{eq:FD_PEEC}
\left[
\begin{array}{cc}
    \boldsymbol{\pazocal Z}(s) + s \mathbf L_p & - \mathbf {A^T} \\
    \mathbf A & s \mathbf P^{-1} + \mathbf{Y}_{\ell e}(s)  
\end{array} 
\right] 
\left[ \begin{array}{c}
       \mathbf I (s)\\ 
       \boldsymbol{\Phi} (s)    
\end{array} \right] = 
\left[ \begin{array}{c}
       \mathbf {V_s}(s)\\
       \mathbf {I_s}(s)
\end{array} \right]
\end{equation}
where $\boldsymbol{\pazocal Z}(s)$ is the impedance matrix accounting for the impedance of conductors or dielectrics elementary volumes, $\mathbf Y_{\ell e}$ is the lumped admittance matrix that contains all the lumped admittances connected to the nodes of the equivalent circuit and $\mathbf A$ is the incidence matrix \cite[Eq. (13.6)]{Ru17AJ}.
\Rob{The knowledge of the voltage and current sources, denoted as $\mathbf V_s(s)$ and $\mathbf I_s(s)$ respectively, allows one to determine the unknown vectors $\boldsymbol{\Phi}(s)$ and $\mathbf I (s)$, which represent the node potentials and branch electric currents, respectively.}
\Rob{Once $\boldsymbol{\Phi}(s)$ and $\mathbf I(s)$ are obtained, $\mathbf{Z}_{sys}$ is determined.}

\Rob{We remark that the described approach yields a comprehensive EM characterization of the considered system, thus providing deep insights on several system features. For instance, the obtained PEEC description can be exploited to accurately characterize the end-to-end channel gain, to design impedance matching strategies and for signal integrity studies. Therefore, the PEEC method represents a versatile and powerful tool for the analysis and the optimization of RIS-aided wireless links. }

\subsection{Analytical Model}
\label{subsec:analytical}
\Rob{According to the methodological approach proposed in \cite{Grad21DR}, the transmitting and receiving antennas, as well as the scattering elements of the RIS, are assumed to be cylindrical thin wire dipoles composed of perfectly conducting material}. These dipoles possess a finite but negligible radius, denoted as $a$, \Rob{which is} much smaller than \Rob{the dipole} length, $\ell$. The thin wire dipoles are connected to complex-valued tunable impedances. 

The \Rob{determination} of the communication channel gain depends on the knowledge of the mutual impedances between the thin wire dipoles. \Rob{Based on \cite{Grad21DR}, the self and mutual impedances describing the system are given by}
%
\begin{equation}\label{eq:Zqp}
        Z_{qp} = \int_{z_q - \ell_q/2}^{z_q + \ell_q/2} \int_{z_p - \ell_p/2}^{z_p + \ell_p/2} g_{qp}(z',z'') \tilde{I}_{z,p}(z') \tilde{I}_{z,q}(z'') dz' dz''
\end{equation}
where $\tilde{I}_{z,\chi}(z') = \text{sin}\left[k_0 \left( \ell_{\chi}/2 - |z'-z_{\chi}| \right) \right] / \text{sin} \left( k_0 \ell_{\chi}/2 \right)$ is the distribution current that is assumed \Rob{to be} sinusoidal, and $ g_{qp}(z',z'') = j\eta_0 (4 \pi k_0)^{-1} \pazocal F_p (\mathbf r_{S_{q}},z') \pazocal G_p (\mathbf r_{S_{q}},z')$. 
The characteristic impedance in vacuum is $\eta_0 = \sqrt{\mu_0 / \varepsilon_0}$, and the wavenumber is $k_0 = 2\pi / \lambda$, where $\mu_0$ and $\varepsilon_0$ are the permittivity and permeability in vacuum, and $\lambda$  is the wavelength. The terms $\pazocal F_p$ and $\pazocal G_p$ \Rob{are defined in \cite{Grad21DR}.}

\section{RIS Optimization}
\label{sec:opt}
The configuration of the RIS is obtained by utilizing the algorithm recently proposed in \cite{HAS23}. The approach is based on the block coordinate descent method, which optimizes the tunable impedances of the RIS iteratively. At each iteration, specifically, the algorithm optimizes a single tunable impedance while keeping the others fixed. The proposed approach is based on the application of Sherman-Morrison’s inversion formula, Sylvester’s determinant theorem, and, Gram-Schmidt’s orthogonalization method. Thanks to this approach, a closed-form solution for the optimal impedance is obtained at each oteration, which ensures that the algorithm requires fewer iterations and less time to converge compared with state-of-the-art benchmarks, while guaranteeing better performance. 

\section{Numerical Results}\label{sec:numres}
\Rob{We consider a scenario consisting of} a transmitter that is constituted by $4$ thin wire dipoles, deployed along the $x$-axis, with the first dipole being centered in $\mathbf r_{Tx} = [0\  0]$ on the $xy$-plane, and the 4 dipoles being spaced by $\lambda /2$; a single receiving dipole of length $\lambda /2$ centered in $\mathbf r_{Rx}=[9.6 \lambda \  14.4 \lambda]$ on the $xy$-plane; and an RIS, which consists of an array of dipoles deployed along the $x$-axis, with the first dipole being centered in $\mathbf r_{RIS}=[0 \  24 \lambda]$. The interdistance of the RIS elements is $d=\lambda /8$ on the $xy$-plane, and three sizes for the RIS are considered with a number of elements $N_{RIS}$ equal to $4$, $16$ and $64$.
%
\begin{figure}[!t]
\centerline{{
\includegraphics[width=1\columnwidth]{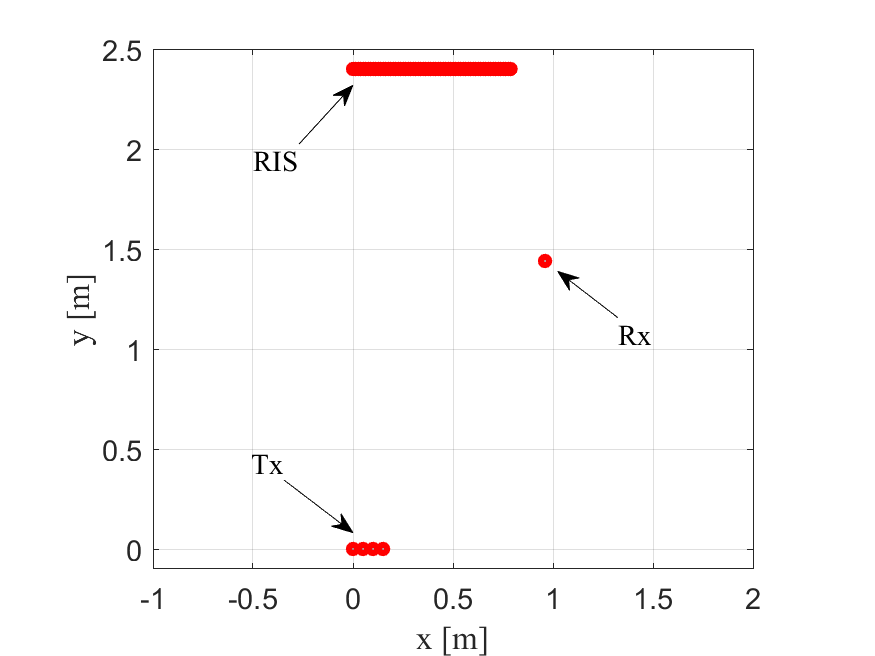}}}
\caption{System setup for a 64 elements RIS.}
\label{fig:ris64}
\end{figure}
Figure \ref{fig:ris64} shows the system configuration of an RIS with $64$ elements.
The resonance frequency is \Rob{set to} $3$ GHz, \Rob{which correspond to a} wavelength of $\lambda=10$ cm.
The \Rob{considered} analytical and numerical \Rob{methods for system characterization are those described in} Section \ref{sec:sysMOD}, and the end-to-end the channel gain is obtained using \eqref{eq:H_E2E}. In \Rob{our evaluation}, the direct path \Rob{between the transmitting and receiving antennas array is assumed to be blocked by obstacles}.
%
\begin{figure}[!t]
\centerline{{
\includegraphics[width=1\columnwidth]{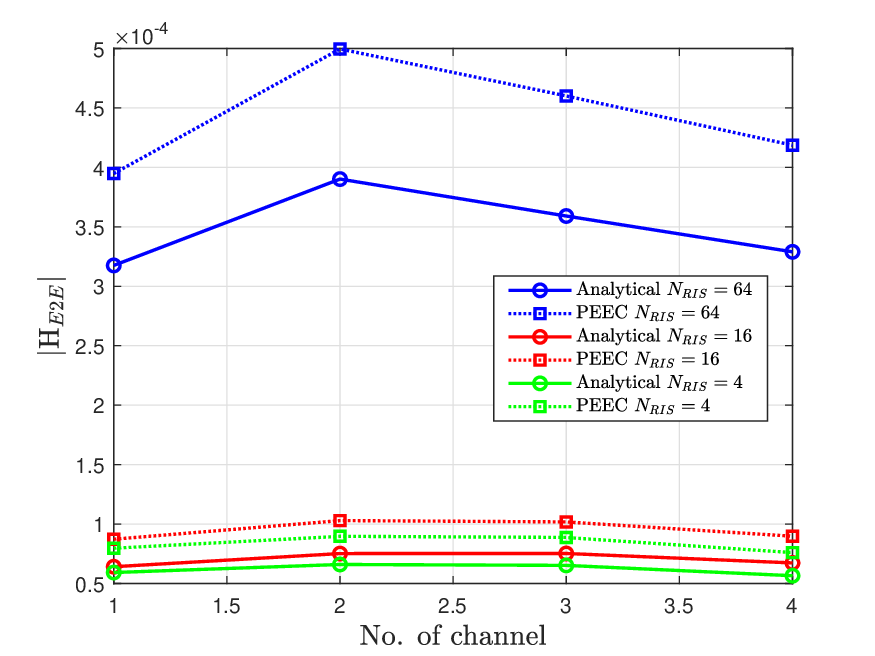}}}
\caption{Channel gain with unoptimized RIS terminations.}
\label{fig:he2e}
\end{figure}

Figure \ref{fig:he2e} illustrates the channel gain when the RIS \Rob{terminations} are not optimized. \Rob{For this evaluation, resistive type RIS terminations are considered, where all the terminations are set to $0.2$ $\Omega$.} \Rob{It is possible to observe that the analytical characterization and the PEEC electromagnetic simulator exhibit consistent performance trends and provide similar performance.}

\begin{figure}[!t]
\centerline{{
\includegraphics[width=1\columnwidth]{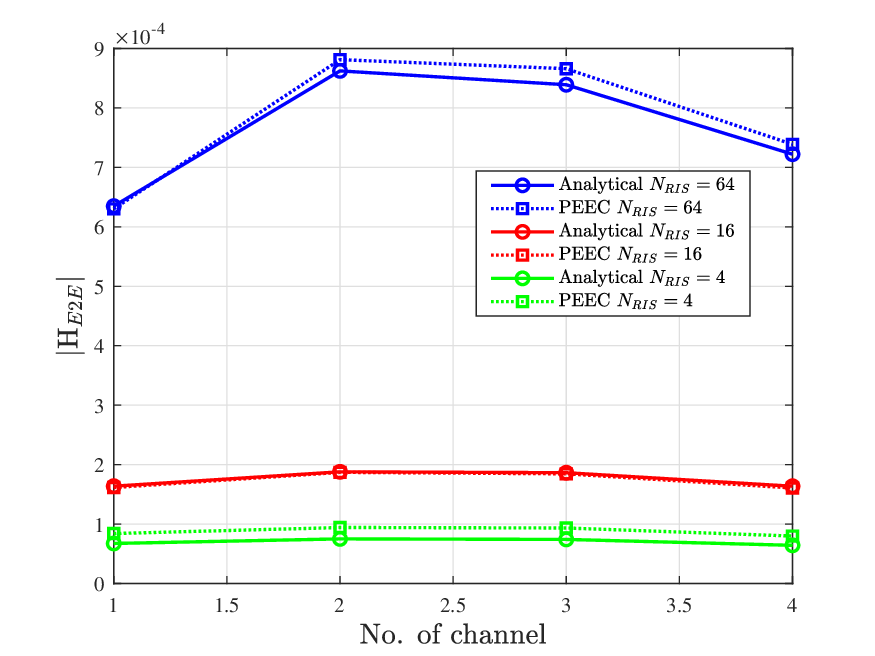}}}
\caption{Channel gain with optimized RIS terminations. }
\label{fig:he2e_opt}
\end{figure}

\Rob{Figure \ref{fig:he2e_opt} compares the channel gain when the RIS terminations are obtained by applying the optimization algorithm described in Section \ref{sec:opt}, where the algorithm is initialized by setting all the terminations to $0.2$ $\Omega$.} It can be observed that optimizing an RIS is a crucial step, since a substantial improvement of the channel gain is obtained. Again, the analytical model and the PEEC simulator are in good agreement.
%
\section{Conclusion}
\label{sec:con}
%
\Rob{In this paper, we validated a recently proposed multiport network model for RIS-assisted wireless communication channels by means of a numerical simulator based on the PEEC method.} 

\section*{ACKNOWLEDGEMENT}
Giuseppe Pettanice acknowledges the Filauro Foundation for supporting him while staying at the Universit\'e Paris-Saclay, CentraleSup\'elec, 91192 Gif-sur-Yvette, France. 
The work of M. Di Renzo was supported in part by the European Commission through the Horizon Europe project titled COVER under grant agreement number 101086228, the Horizon Europe project titled UNITE under grant agreement number 101129618, and the Horizon Europe project titled INSTINCT under grant agreement number 101139161, as well as by the Agence Nationale de la Recherche through the France 2030 project titled ANR-PEPR Networks of the Future under grant agreement NF-PERSEUS, 22-PEFT-004, and by the CHIST-ERA project titled PASSIONATE under grant agreement CHIST-ERA-22-WAI-04 through ANR-23-CHR4-0003-01.
This work was also partially supported by the Centre of Excellence on Connected, Geo-Localized, and Cyber Secure Vehicles (Ex-EMERGE), funded by the Italian Government under CIPE Resolution 70/2017. 

\bibliographystyle{IEEEtran}
\bibliography{bibl/Abbr12, bibl/Ref2, bibl/Ref14}

\end{document}